\documentclass[5p,twocolumn,10pt]{elsarticle}
\usepackage{rotate}
\usepackage{lscape}
\usepackage{epsfig}
\usepackage{hyperref}
\usepackage{amsmath}
\usepackage{multirow}
\usepackage{subfig}

\biboptions{sort&compress}

\newcommand{\bq}{\begin{equation}}
\newcommand{\eq}{\end{equation}}
\newcommand{\bqa}{\begin{eqnarray}}
\newcommand{\eqa}{\end{eqnarray}}
\newcommand{\baa}[1]{\begin{array}{#1}}
\newcommand{\eaa}{\end{array}}

\def\MSbar{$\overline{\mathrm{MS}}\ $}

\def\gsim{\mathrel{\raise.3ex\hbox{$>$\kern-.75em\lower1ex\hbox{$\sim$}}}}
\def\lsim{\mathrel{\raise.3ex\hbox{$<$\kern-.75em\lower1ex\hbox{$\sim$}}}}

\newcounter{bla}

\journal{Computer Physics Communications}

\begin{document}

\begin{frontmatter}

\title{NLO EW and QCD proton-proton cross section calculations with \texttt{mcsanc}-v1.01}


\author[a,b]{Sergey~G.~Bondarenko}
\author[a]{Andrey~A.~Sapronov\corref{author}}

\cortext[author] {Corresponding author.\\\textit{E-mail address:} andrey.a.sapronov@gmail.com}
\address[a]{Dzhelepov Laboratory for Nuclear Problems, JINR, Joliot-Curie 6, RU-141980 Dubna, Russia}
\address[b]{Bogoliubov Laboratory of  Theoretical Physics, JINR, Joliot-Curie 6, RU-141980 Dubna, Russia}


\begin{abstract}
\texttt{mcsanc} is a Monte Carlo tool based on the SANC (Support for Analytic
and Numeric Calculations for experiments at colliders) modules for higher order
calculations in hadron collider physics. It allows to evaluate NLO QCD and EW
cross sections for Drell--Yan processes (inclusive), associated Higgs and gauge
boson production and single-top quark production in \(s\)- and \(t\)-channel. The paper
contains theoretical description of the SANC approach, numerical validations
and manual.
\end{abstract}

\begin{keyword}
Perturbation theory; NLO calculations; Standard Model; Electroweak interaction; QCD; QED; Monte Carlo integration;

\end{keyword}

\end{frontmatter}


{\bf PROGRAM SUMMARY}

\begin{small}
\noindent
{\em Program Title: mcsanc-v1.01}                              \\
{\em Journal Reference:}                                      \\
{\em Catalogue identifier:}                                   \\
{\em Licensing provisions: none}                              \\
{\em Programming language: Fortran, C, C++}                   \\
{\em Computer: x86(-64) architecture}                                 \\
{\em Operating system: Linux}                                 \\
{\em RAM:} 2 Gbytes                                           \\
{\em Number of processors used: multiprocess}                 \\
{\em Classification:11.1, 11.6}                               \\
{\em External routines/libraries: LHAPDF~[1]}                     \\
{\em Subprograms used: Looptools~[2], Cuba~[3]}              \\
{\em Nature of problem: Theoretical calculations at next-to-leading order in
perturbation theory allow to compute higher precision amplitudes for Standard
Model processes and decays, provided proper treatments of UV divergences and IR
singularities are performed.}\\
   \\
{\em Solution method: Numerical integration of the precomputed differential expressions for partonic 
cross sections of certain processes implemented as SANC modules~[4]. }\\
   \\
{\em Restrictions: the list of processes is limited to Drell--Yan, associated Higgs and 
gauge boson production and single-top production in \(s\)- and \(t\)-channel}\\
   \\
{\em Running time: from hours to days depending on requested precision and kinematic conditions}\\
   \\

\end{small}

\section{Introduction \label{intro}}

Most demanded theoretical calculations for high energy particle interaction
processes find their realization as a Monte Carlo code usable by
experimentalists.
The Monte Carlo tools can fulfill several tasks: calculate total and
differential cross section or provide events in final state kinematics suitable
for analysis simulation. For the first type we use here the term
``integrator''.  The second kind of tools is known as generators.  Besides the
calculation itself, such qualities as speed, functionality and configuration
simplicity are valued.

The fixed order calculations for the Drell--Yan process are implemented in
several Monte Carlo tools, providing evaluations for fully differential
inclusive cross sections. At the NNLO QCD level there are
DYNNLO~\cite{Catani:2007vq, Catani:2009sm} and FEWZ~\cite{Anastasiou:2003ds,
Gavin:2010az} integrators. Recent FEWZ\_v3.0~\cite{Li:2012wna} also includes
electroweak NLO corrections.  The DYRAD package provides NLO QCD calculations
in the context of jets production on colliders utilizing the method of
universal crossing functions~\cite{Giele:1993dj}.  The well known
MCFM~\cite{Campbell:1999ah} package includes NLO QCD calculations for DY,
associated Higgs and gauge boson production, single top quark production and
large amount of other processes predicted for proton-(anti)proton collisions.
There are number of codes that compute complete \(\mathcal{O}(\alpha)\)
corrections. There are WGRAD/WGRAD2~\cite{Baur:1998kt, Baur:2004ig} and
ZGRAD2~\cite{Baur:2001ze} for Drell--Yan charged and neutral currents
respectively and the HORACE~\cite{CarloniCalame:2007cd, CarloniCalame:2006zq}
code for both neutral and charged currents.  Besides the mentioned software
realizations there is a large number of theoretical calculations for complete higher
order QED and electroweak~\cite{Mosolov:1981xk, Soroko:1990ug,
Wackeroth:1996hz, Baur:1997wa, Dittmaier:2001ay, Zykunov:2001mn,
Brensing:2007qm, Zykunov:2005tc, Dittmaier:2009cr, Arbuzov:2005dd,
Arbuzov:2007ke} and QCD~\cite{Aurenche:1980tp, Hamberg:1990np,
vanNeerven:1991gh, Melnikov:2006di, Andonov:2009nn} corrections to the DY
processes.

A combined treatment of QCD and EW corrections is being developed in other
studies, besides the mentioned FEWZ\_v3.0 codes. The POWHEG BOX tool includes
exact NLO EW and QCD corrections to the DY CC~\cite{Bernaciak:2012hj,
Barze:2012tt} and NC~\cite{Barze':2013yca} processes in a single computational
framework.

The NLO QCD corrections to \(pp\to HZ(W)\) processes, similar to those for the
DY, are presented in~\cite{Han:1991ia, Ciccolini:2003jy} and NNLO QCD ones are
calculated in~\cite{Brein:2003wg}. Electroweak corrections to the associated
Higgs and gauge boson production are discussed in~\cite{Ciccolini:2003jy} 
and~\cite{Bardin:2005dp}.

The single-top production in \(s\)- and \(t\)-channel at hadron colliders is
considered in the context of NLO QCD in~\cite{Smith:1996ij, Bordes:1994ki,
Stelzer:1997ns, Stelzer:1998ni, Harris:2002md, Sullivan:2004ie,
Campbell:2004ch}. NLO EW contiributions are presented
in~\cite{Beccaria:2006ir, Beccaria:2008av} (including MSSM extension) and
in~\cite{Bardin:2010mz}.

In this paper we describe a Monte Carlo integrator
\texttt{mcsanc}-v1.01~\cite{Bardin:2012jk} in which the SM processes of DY,
Higgs-strahlung and single-top production in \(s\)- and \(t\)-channel are implemented
with NLO QCD and EW corrections.

The paper is organized as follows. Section~\ref{sec:ho-sanc} contains details
on next to leading order calculations in the SANC system, it describes
treatment of singularities, electroweak scheme issues and code organization.
Section~\ref{sec:num-val} gives overview of numerical cross checks and validation
performed for \texttt{mcsanc}-v1.01 program and summary is given in 
Section~\ref{sec:summary}. \ref{sec:mcsanc-prog} contains
technical information on the program's usage and configuration and \ref{sec:multiproc-eff}
describes efficiency issues related to multiprocess runs.

\section{NLO corrections in SANC \label{sec:ho-sanc}}


The SANC system~\cite{Andonov:2004hi,Andonov:2008ga} implements calculations of
complete (real and virtual) NLO QCD and EW corrections for the Drell--Yan
process, associative Higgs and gauge boson production, single top production
and several other processes at partonic level.  Here we give a brief summary of
the main properties of this framework. For complete list of SANC processes
see~\cite{Andonov:2008ga}.

All calculations are performed within the OMS (on-mass-shell) renormalization
scheme \cite{Bardin:1999ak} in the $R_\xi$ gauge which allows an explicit control
of the gauge invariance by examining a cancellation of the gauge parameters in the
analytical expression of the matrix element.

Depending on the process and type of corrections, we subdivide the total 
NLO cross section at the partonic level into several terms:
\begin{align}
d\sigma = & d\sigma^{\mathrm{LO}}+
d\sigma^{\mathrm{virt}}(\lambda) \nonumber \\
& +d\sigma^{\mathrm{soft}}(\lambda,\bar{\omega})
+d\sigma^{\mathrm{hard}}(\bar{\omega}),
\end{align}
differential over a
generic observable which is a function of the final state momenta,
$\sigma^{\mathrm{LO}}$ is the leading order cross section,
$\sigma^{\mathrm{virt}}$ is a contribution of virtual (loop) corrections,
$\sigma^{\mathrm{soft}}$ corresponds to a soft photon or gluon emission and
$\sigma^{\mathrm{hard}}$ is a contribution of a hard (real) photon or gluon emission. The
terms with auxiliary parameters $\bar{\omega}$ (photon energy which separates
phase spaces associated with the soft and hard photon emission) and
$\lambda$ (photon mass which regularizes infrared divergences) are introduced 
in the NLO EW calculations. They cancel out after
summation and the differential NLO EW cross-section for infrared-safe
observables does not depend on these parameters~\cite{Greco:1980mh,Bohm:1982hr,Denner:1991kt}.

\begin{figure}
\begin{center}
  \includegraphics[width=0.45\textwidth]{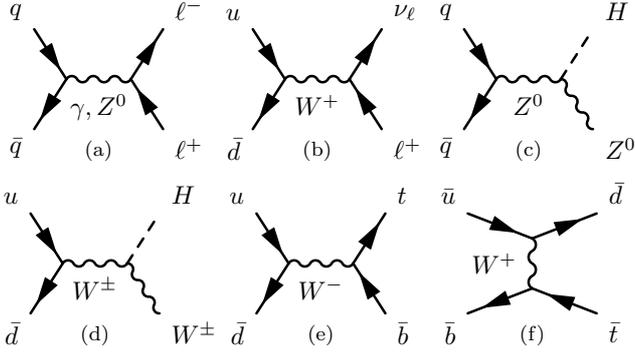}
\end{center}
\caption{Feynman graphs for tree level Drell-Yan process
neutral~(\textit{a}) and charged~(\textit{b}) currents,
Higgs and gauge boson production neutral~(\textit{c}) and
charged~(\textit{d}) currents, and single top-quark production
\(s\)-channel~(\textit{e}) and \(t\)-channel~(\textit{f}).}
\label{fig:fdiag-lo}
\end{figure}

The list of processes implemented in the \texttt{mcsanc}-v1.01 code is given in the Table~\ref{tab:mcsanc-proc-list}
with the tree level diagrams shown in
Figure~\ref{fig:fdiag-lo}.

\begin{table}
\begin{center}
\begin{tabular}{|r|l|c|}
\hline
pid & \(f(p_1)+f(p_2)\to\) & ref.\\
\hline
001 & \( e^+(p_3) + e^-(p_4)\)                      & \multirow{3}{*}{\cite{Arbuzov:2007db,Andonov:2009nn}}\\
002 & \( \mu^+(p_3) + \mu^-(p_4)\)                  & \\
003 & \( \tau^+(p_3) + \tau^-(p_4)\)                & \\
\hline
004 & \( Z^0(p_3) + H(p_4)\)                        & \cite{Bardin:2005dp}  \\
\hline
101 & \( e^+(p_3) + {\nu}_e(p_4)\)                 & \multirow{6}{*}{\cite{Arbuzov:2005dd,Andonov:2009nn}}\\
102 & \( \mu^+(p_3) + {\nu}_{\mu}(p_4)\)           & \\
103 & \( \tau^+(p_3) + {\nu}_{\tau}(p_4)\)         & \\
-101 & \( e^-(p_3) + \bar{\nu}_e(p_4)\)            & \\
-102 & \( \mu^-(p_3) + \bar{\nu}_{\mu}(p_4)\)      & \\
-103 & \( \tau^-(p_3) + \bar{\nu}_{\tau}(p_4)\)    & \\
\hline
104 & \( W^+(p_3) + H(p_4)\)                       & \multirow{2}{*}{-}\\
-104 & \( W^-(p_3) + H(p_4)\)                      & \\
\hline
105 & \( t(p_3) + \bar{b}(p_4)\) (\(s\)-channel)       & \multirow{4}{*}{\cite{Bardin:2010mz,Bardin:2011ti}}\\
106 & \( t(p_3) + q(p_4)\) (\(t\)-channel)             & \\
-105 & \( \bar{t}(p_3) + {b}(p_4)\) (\(s\)-channel)    & \\
-106 & \( \bar{t}(p_3) + q(p_4)\) (\(t\)-channel)      & \\
\hline
\end{tabular}
\caption{List of the processes implemented in the \texttt{mcsanc}-v1.01 integrator with references.} \label{tab:mcsanc-proc-list}
\end{center}
\end{table}

\begin{figure}
\begin{center}
  \includegraphics[width=0.45\textwidth]{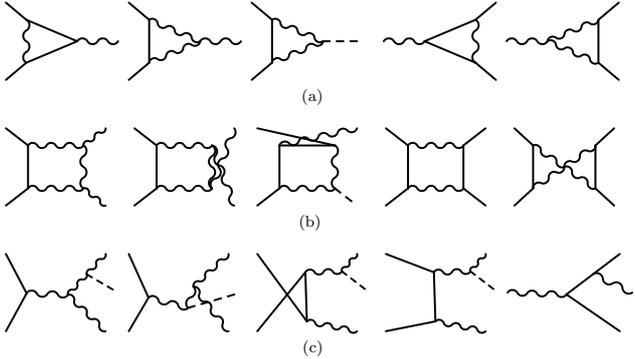}
\end{center}
\caption{Examples of Feynman graphs contributing to EW NLO corrections for Drell-Yan process 
and associated Higgs and gauge boson production: (\textit{a})~vertices, 
(\textit{b})~boxes and (\textit{c})~emission.}
\label{fig:fdiag-ewnlo}
\end{figure}

\begin{figure}
\begin{center}
  \includegraphics[width=0.45\textwidth]{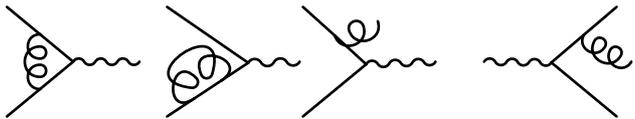}
\end{center}
\caption{Examples of Feynman graphs contributing to QCD NLO corrections in \texttt{mcsanc}-v1.01 integrator.}
\label{fig:fdiag-qcdnlo}
\end{figure}

Figures~\ref{fig:fdiag-ewnlo} and~\ref{fig:fdiag-qcdnlo} show examples of 
diagrams of corresponding NLO EW and QCD contributions to the implemented processes.

Electroweak NLO radiative corrections contain terms proportional to logarithms of
the quark masses, $\ln(\hat{s}/m_{u,d}^2)$.  They come from the initial state
radiation contributions including hard, soft and virtual photon emission. Such
initial state mass singularities are well known, for instance, in the process
of $e^+e^-$ annihilation. However, in the case of hadron collisions these logarithms have
been already {\em effectively} taken into account in the parton density
functions (PDF).  In fact, in the procedure of PDF extraction from 
experimental data, the QED radiative corrections to the quark line were not
systematically subtracted. Therefore current PDFs effectively include not
only the QCD evolution but also the QED one.  Moreover, it is known that the
leading logarithmic behaviours of the QED and QCD DGLAP evolution of the quark density
functions are similar (proportional to each other). Consequently one gets an evolution of
the PDF with an effective coupling constant
\begin{align}
\alpha^{\mathrm{eff}}_{s} \approx \alpha_{s} + \frac{Q_i^2}{C_F}\alpha,
\end{align}
where $\alpha_s$ is the strong coupling constant, $\alpha$ is the fine
structure constant, $Q_i$ is the quark charge, and $C_F$ is the QCD colour
factor.

The system supports both \MSbar and DIS subtraction schemes. 
A solution described in \cite{Diener:2005me} allows to avoid 
the double counting of the initial quark mass
singularities contained in the results for the corrections to the free quark
cross section and the ones contained in the corresponding PDF.  The latter
should also be taken in the same scheme with the same factorization scale.

For example, the \MSbar subtraction to the fixed (leading) order in $\alpha$ is given by:
\begin{align}
\label{msbarq}
\begin{split}
\bar{q}(x,M^2)& = q(x,M^2) -
\int_x^1 \frac{\mathrm{d} z}{z} \, q\biggl(\frac{x}{z},M^2\biggr) \,
\frac{\alpha}{2\pi} \, Q_q^2
\nonumber \\ & \quad \times
\biggl[ \frac{1+z^2}{1-z}
\biggl\{\ln\biggl(\frac{M^2}{m_q^2}\biggr)-2\ln(1-z)-1\biggr\} \biggr]_+
\\ 
& \equiv q(x,M^2) - \Delta q,
\end{split}
\end{align}
where \(q(x,M^2)\) is the parton density function in the \MSbar scheme computed
using the QED DGLAP evolution.

The differential hadronic cross section for DY processes (\(001\div 003, \pm101\div 103\))
is given by
\begin{align}
  \mathrm{d}\sigma^{pp \to \ell\ell'X} = & \sum_{q_{1}q_{2}} \int\limits_0^1 \int\limits_0^1 
  \mathrm{d}x_1 \, \mathrm{d}x_2 \, \bar{q}_1(x_1,M^2) \, \bar{q}_2(x_2,M^2)\,\nonumber \\
  & \times \mathrm{d}\hat{\sigma}^{q_1 q_2\to \ell\ell'},
\end{align}
where \(\bar{q}_1(x_1,M^2), \bar{q}_2(x_2,M^2) \) are the parton density
functions of the incoming quarks modified by the subtraction of the quark mass
singularities and \( \sigma^{q_1 q_2\to \ell\ell'} \) is the partonic cross
section of corresponding hard process.  The sum is performed over all
possible quark combinations for a given type of process ($q_1q_2 = u\bar{d},
u\bar{s}, c\bar{d}, c\bar{s}$ for CC and $q_1q_2 = u\bar{u}, d\bar{d},
s\bar{s}, c\bar{c}, b\bar{b}$ for NC). The expressions for other processes are similar.

The effect of applying different EW schemes in the SANC system is discussed
in~\cite{Arbuzov:2007db}. The SANC system supports \(\alpha(0), G_{\mu}, \alpha(M_Z)\) (DY NC only) 
schemes (see comments on EW parameters in~\ref{sec:mcsanc-prog}), of which
the \(G_{\mu}\)-scheme~\cite{Degrassi:1996ps} can be preferable since it minimizes EW radiative
corrections to the inclusive DY and Higgs-strahlung cross section.  In this scheme the weak coupling
\(g\) is related to the Fermi constant and the W boson mass by equation
\begin{align}
g^2 = 4\sqrt{2} G_{\mu} m_{W}^2(1-\Delta r),
\label{eqn:coupling}
\end{align}
where \(\Delta r\) represents all radiative corrections to the muon decay
amplitude~\cite{Sirlin:1980nh}. Since the vertex term between charged particles and
photons is proportional to \(g \sin{\theta_W}\), one can introduce an effective
electromagnetic coupling constant
\begin{align}
\alpha_{G_{\mu}}^{tree} = \frac{\sqrt{2} G_{\mu} \sin^2\theta_W m_W^2}{\pi},
\end{align}
which is evaluated from~(\ref{eqn:coupling}) in a tree-level approximation by
setting \(\Delta r = 0 \). 

\begin{figure}[h]
\begin{center}
  \includegraphics[width=0.45\textwidth]{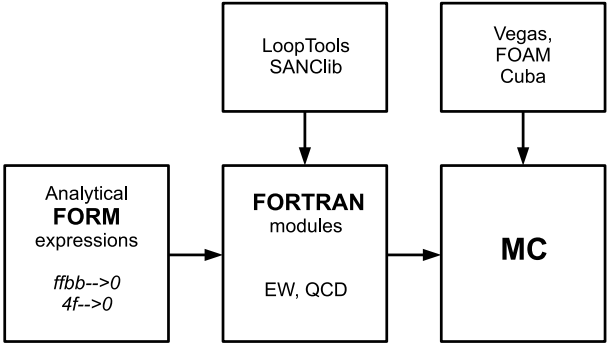}
\end{center}
\caption{The SANC framework scheme.}
\label{fig:sanc-scheme}
\end{figure}

The scheme of the SANC framework is shown on the Figure~\ref{fig:sanc-scheme}. Analytical
expressions are obtained for the formfactors and amplitudes
of generalized processes \(ffbb \to 0\) and \(4f\to 0\) and stored as the FORM~\cite{Vermaseren:2000nd}
language expressions~\cite{Andonov:2004hi,Bardin:2005dp,Arbuzov:2007ke}. The latter are translated to the Fortran modules~\cite{Andonov:2008ga} for
specific parton level processes with NLO QCD and EW corrections. The modules
are utilising Looptools~\cite{Hahn:1998yk} and SANClib~\cite{Bardin:2009ix} packages for loop integrals evaluation. 
To build a Monte Carlo code one convolutes the partonic cross sections from
the modules with the parton density functions and feeds the result as an 
integrand to any Monte Carlo algorithm implementation, e.g. FOAM~\cite{Jadach:2005ex} or Cuba~\cite{Hahn:2004fe}.
The module's procedures for partonic cross sections are significantly unified and allow
to calculate fully differential hadronic cross sections.

\section{Numerical validation \label{sec:num-val}}

The numerical validation was performed using setup given below.

\begin{tabular}{ll}
$\alpha =1/137.035999679$,      & $G_{F} = 1.16637\times 10^{-5}$,  \\
$M_{W}= 80.399$\,GeV,           & $M_{Z}= 91.1876$\,GeV,            \\
$M_{H}= 120$\,GeV,              &                                   \\
$\Gamma_Z$ = 2.4952\,GeV,       & $\Gamma_W$ = 2.085\,GeV, 	    \\
$V_{ud} = 0.9738$, 		& $V_{us}= 0.2272$, 		    \\
$V_{cd}= 0.2271$,               & $V_{cs}= 0.9730$.		    \\
$V_{ub}= 0.$,		        & $V_{cb}= 0.$,                     \\
$m_e = 0.510998910$\,MeV,	& $m_\mu = 0.105658367$\,GeV,       \\
$m_\tau = 1.77682$\,GeV,        &				    \\
$m_u = 0.066$\,GeV,		& $m_d = 0.066$\,GeV,		    \\
$m_c = 1.2$\,GeV,		& $m_s = 0.150$\,GeV,		    \\
$m_t = 172.9$\,GeV,             & $m_b = 4.67 $\,GeV. 		    \\
\end{tabular}
The electroweak parameters in the \(G_{\mu}\) scheme were taken from 
PDG-2011~\cite{Nakamura:2010zzi}, except for
the light quark masses which were taken from~\cite{Dittmaier:2009cr}.
The PDF set was CT10~\cite{Guzzi:2011sv} interfaced via LHAPDF~\cite{Whalley:2005nh} 
library with the factorization scale equal to renormalization
scale and particular for the processes under consideration: $M_V$ for DY-like single $V$ production;
$M_{V+H}$ for the processes \(004, \pm104\); $m_{t}$ for the processes \(\pm105, \pm106\).

The phase-space cuts were set to loose values: for the final state particle transverse momenta $p_{T}\geq 0.1\,$GeV,
no cuts for their rapidities and for the neutral current DY, in addition, $M_{l^+l^-}\geq 20\,$GeV.
The numerical results are provided for DY process only for muon case ignoring the effects of
recombination. The  \(\bar{\omega}\) parameter was set to = \(10^{-4}\) and the CMS 
energy to $\sqrt{s_0}=14\,$TeV if not stated otherwise.

At NLO level several hard sub-processes contribute to a given process. In
general, it consists of several parts: LO -- leading order,
\textit{virt} -- virtual, \textit{real} brems(glue)-strahlung and \textit{subt} --
subtraction; \textit{real}, in turn, is subdivided into \textit{soft} and \textit{hard} contributions. 
All contributions are enumerated with {\tt id=0}--{\tt 6}:
\begin{itemize}
\item[{\tt id0}:] LO, \(2\to 2\), tree-level, $q\bar{q'}$ NC or CC sub-processes.
\item[{\tt id1}:] \textit{subt} term, responsible for the subtraction of the initial 
	   quark mass ($m_q$) singularities for $q\bar{q'}$ sub-processes,
	   computed in a given subtraction scheme ($\overline{\mbox{MS}}$ or
	   DIS). It depends on $\ln(m_q)$.
\item[{\tt id2}:] \textit{virt} represents only the NLO EW parts, stands for pure EW 
	   one-loop virtual contributions. It depends on $m_q$ and may depend
	   on an infrared regulator (e.g. on an infinitesimal photon mass). It
	   is not present for NLO QCD contributions, where it is added to the
	   soft contribution (see next item).  For DY NC NLO EW process it
	   contains all virtual contributions, both EW and QED.
\item[{\tt id3}:] For all processes, except DY NC, this stands for the sum of \textit{virt} 
	   and \textit{real} \textit{soft} (QED/QCD) contributions, therefore
	   it does not depend on the infrared regulator but depends on $m_q$
	   and on the soft-hard separator $\bar{\omega}$.  For DY NC NLO EW
	   processes it is just the \textit{real} \textit{soft} QED
	   contribution that depends on the infrared regulator, on $m_q$ and on
	   the soft-hard separator $\bar{\omega}$.
\item[{\tt id4}:] For all processes this is just the \textit{real} \textit{hard} (QED/QCD)
           contribution that depends on $m_q$ and on $\bar{\omega}$.
\item[{\tt id5}:] \textit{subt} term is responsible for the subtraction of the initial 
	   quark mass singularities for $gq(gq')$ sub-processes (also
	   computed in $\overline{\mbox{MS}}$ or DIS schemes). It contains
	   logarithmic singularities in $m_q$.
\item[{\tt id6}:] The gluon-induced sub-process -- an analog of {\tt id4} for $gq(gq')$ 
	   sub-processes. They also contain logarithmic mass singularities
	   which cancel those from {\tt id5}.

\end{itemize}

The quark mass is used to regularize the collinear divergences, the soft-hard
separator is a remainder of infrared divergences.  The sum of contributions
with {\tt id3} and {\tt id4} is independent of $\bar{\omega}$.  The sums {\tt
id1+id2+id3+id4} and {\tt id5+id6} are separately independent of $m_q$.
Therefore, the entire NLO sub-process is independent of both unphysical
parameters $\bar{\omega}$ and $m_q$.
\begin{table}
\centering
\begin{tabular}{|l|r|r|r|}
\hline
 \(pp\to\)       & \(Z^0(\mu^+\mu^-)\) &   \(W^+(\mu^+\nu_{\mu})\)  &   \(W^-(\mu^-\bar{\nu}_{\mu})\)      \\ \hline 
 LO              &   3338(1)   &   10696(1)   &   7981(1)   \\
 LO MCFM         &   3338(1)   &   10696(1)   &   7981(1)   \\ \hline
 NLO QCD         &   3388(2)   &   12263(4)   &   9045(4)   \\ 
 NLO MCFM        &   3382(1)   &   12260(1)   &   9041(5)   \\ \hline
 NLO EW          &   3345(1)   &   10564(1)   &   7861(1)   \\ \hline
$\delta_{QCD}$, \%    &   1.49(3)   &   14.66(1)   &   13.35(3)  \\
$\delta_{EW}$, \%     &   0.22(1)   &   -1.23(1)   &   -1.49(1)  \\  \hline
\end{tabular}
\caption{NC and CC DY processes, i.e. for pid $ = 002, \pm 102$.
LO, NLO EW, NLO QCD cross sections are given in picobarns and compared
with corresponding values obtained with the program MCFM. The
correction factors \(\delta\) are shown in \%.} \label{tab:dy-total-xs}
\end{table}
\begin{table}[h]
\centering
\begin{tabular}{|l|r|r|r|}
\hline
 \(pp\to\)       &  \(Z^0+H\)  &  \(W^++H\)   &   \(W^-+H\)   \\ \hline
 LO              &  0.8291(1)  &  0.9277(1)   &   0.5883(1)   \\
 LO MCFM         &  0.8292(1)  &  0.9280(2)   &   0.5885(1)   \\ \hline
 NLO QCD         &  0.9685(3)  &  1.0897(3)   &   0.6866(3)   \\
 NLO MCFM        &  0.9686(1)  &  1.0901(2)   &   0.6870(1)   \\ \hline
 NLO EW          &  0.7877(1)  &  0.8672(2)   &   0.5508(1)   \\ \hline
$\delta_{QCD}, \%$    &  16.81(3)   &  17.47(3)    &   16.72(5)    \\
$\delta_{EW},\%$     & -5.00(2)    & -6.52(2)     &  -6.38(3)     \\  \hline
\end{tabular}
\caption{The same as in Table \ref{tab:dy-total-xs} but for
processes of $HZ(W^{\pm})$ production, i.e. pid$ = 004, \pm 104$.} \label{tab:hv-total-xs}
\end{table}
\begin{table}
\centering
\begin{tabular}{|l|c|c|}
\hline
\(pp\to\)        &   \(t+\bar{b}\)     &   \(\bar{t}+b\)    \\
                 &   (\(s\)-channel)       &   (\(s\)-channel)      \\ \hline
LO               &   5.134(1)          &   3.205(1)         \\
LO MCFM          &   5.133(1)          &   3.203(1)         \\ 
NLO QCD          &   6.921(2)          &   4.313(2)         \\           
NLO MCFM         &   6.923(2)          &   4.309(1)         \\ 
NLO EW           &   5.022(1)          &   3.140(1)         \\ \hline
$\delta_{QCD},\%$    &   34.79(5)          &   34.56(8)     \\
$\delta_{EW},\%$     &  -2.18(1)           &  -2.02(2)      \\  
\hline 
\hline 
\(pp\to\)        &   \(t+q\)         &   \(\bar{t}+q\)   \\
                 &   (\(t\)-channel)       &   (\(t\)-channel)     \\ \hline
LO               &   158.73(2)         &   95.18(2)        \\
LO MCFM          &   158.69(7)         &   95.27(4)        \\ 
NLO QCD          &   152.13(9)         &   90.44(7)        \\           
NLO MCFM         &   152.07(14)        &   90.50(8)        \\ 
NLO EW           &   164.44(5)         &   98.65(4)        \\ \hline
$\delta_{QCD},\%$    &  -4.17(6)         &   -4.08(8)        \\
$\delta_{EW},\%$     &   3.59(3)         &    3.66(5)        \\  
\hline 
\end{tabular}
\caption{The same as in Table \ref{tab:dy-total-xs} but for single top production, \(s\)- and \(t\)-channel,
i.e. for pid$ = \pm 105,\pm 106$.} \label{tab:st-total-xs}
\end{table}

Below we provide numerical cross checks for the \texttt{mcsanc-v1.01}
integrator. The SANC DY NLO electroweak corrections were thoroughly compared
with other calculations earlier in~\cite{Buttar:2006zd, Gerber:2007xk,
Buttar:2008jx} during theoretical workshops on the subject. Therefore they are
not repeated here. The newer QCD results are validated using the MCFM
program~\cite{Campbell:2010ff}.  At the moment this work is being completed the
comparison of Monte Carlo codes for the accurate description of the Drell--Yan
processes at hadron colliders is ongoing within the \(W\)-mass workshop held by
the EW working group of the LPCC~\cite{ew-lpcc:www}. The resulting publication will
contain a detailed cross check of several MC instruments including
\texttt{mcsanc}-v1.01.

\begin{figure}[h]
  \includegraphics[width=0.5\textwidth]{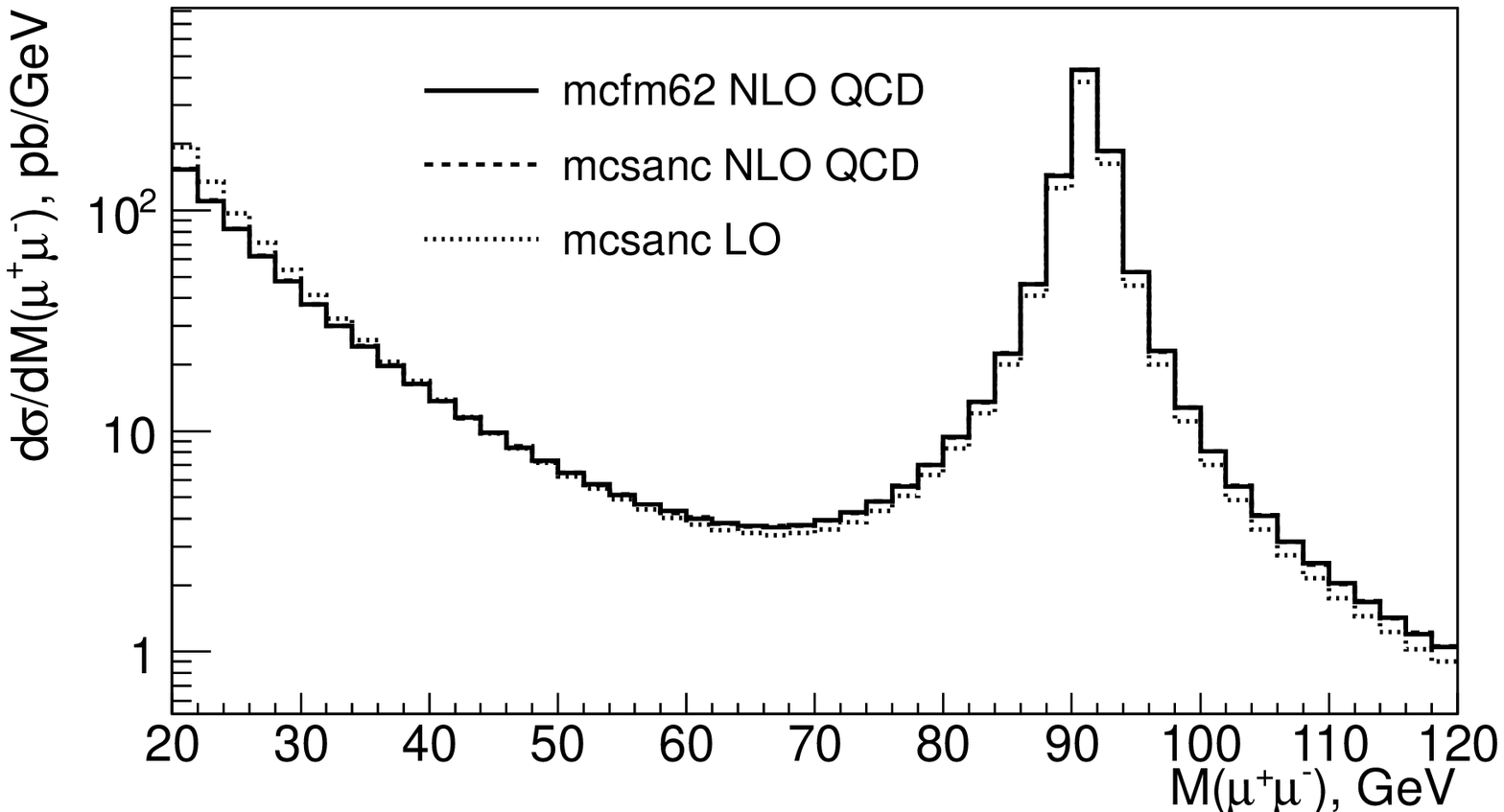}
  \includegraphics[width=0.5\textwidth]{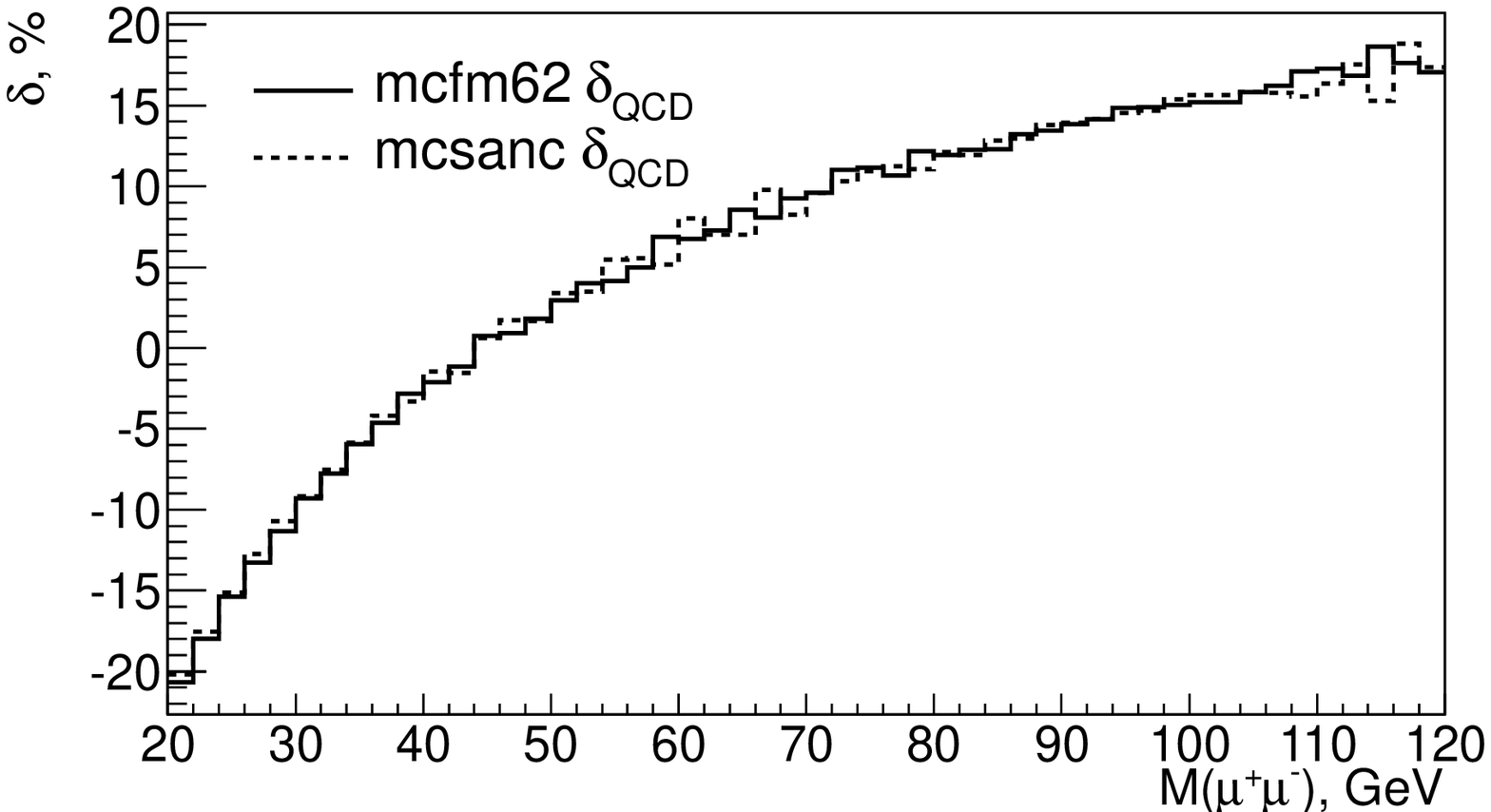}
\caption{Comparison of differential cross sections and correction factors
\(\delta(\mathrm{QCD})\) for neutral current Drell--Yan \(pp\to \mu^+\mu^-\)
process in dimuon invariant mass distribution.}
\label{fig:nc-vs-mcfm-minv}
\end{figure}

\begin{figure}[h]
  \includegraphics[width=0.5\textwidth]{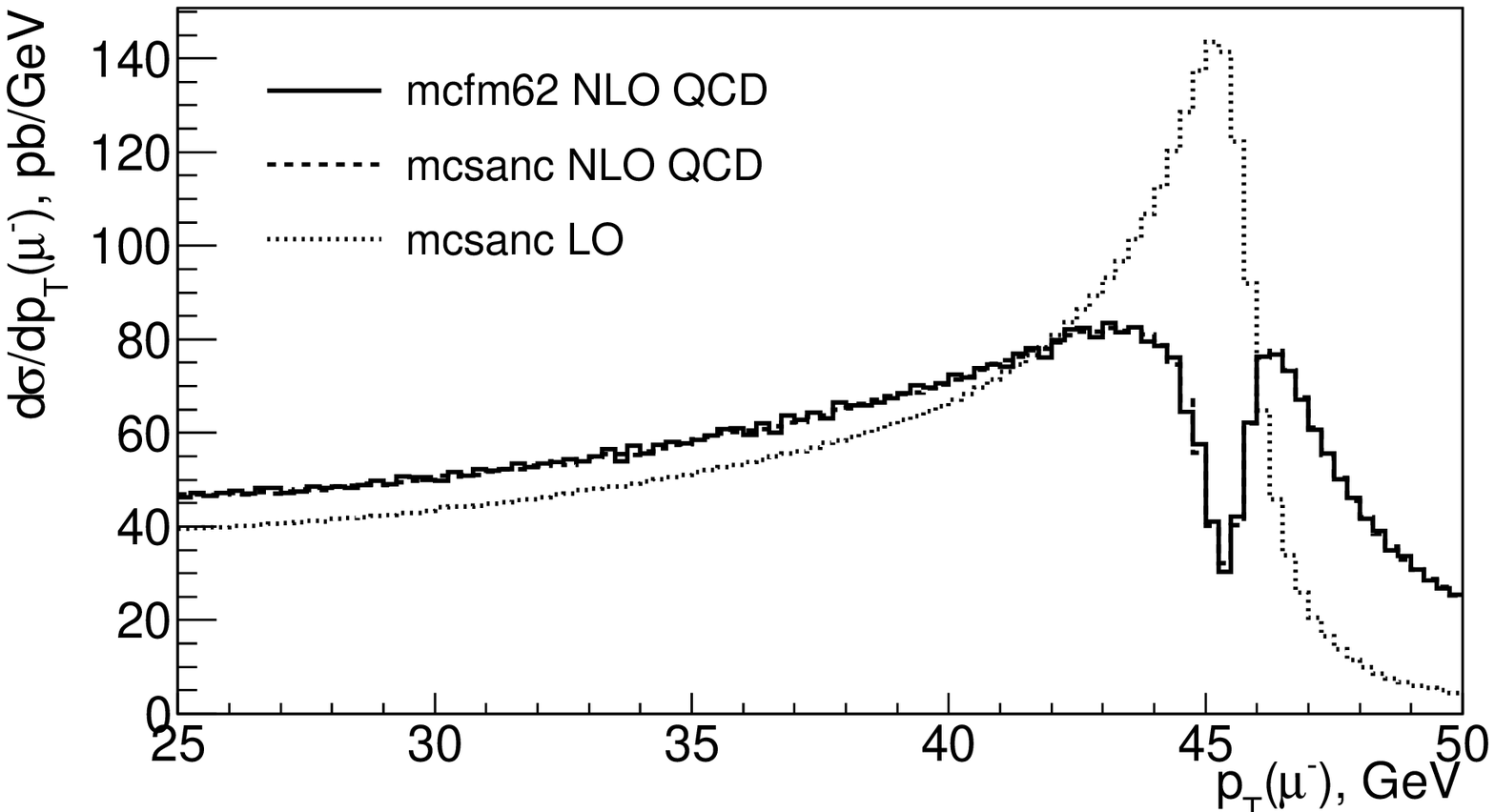}
  \includegraphics[width=0.5\textwidth]{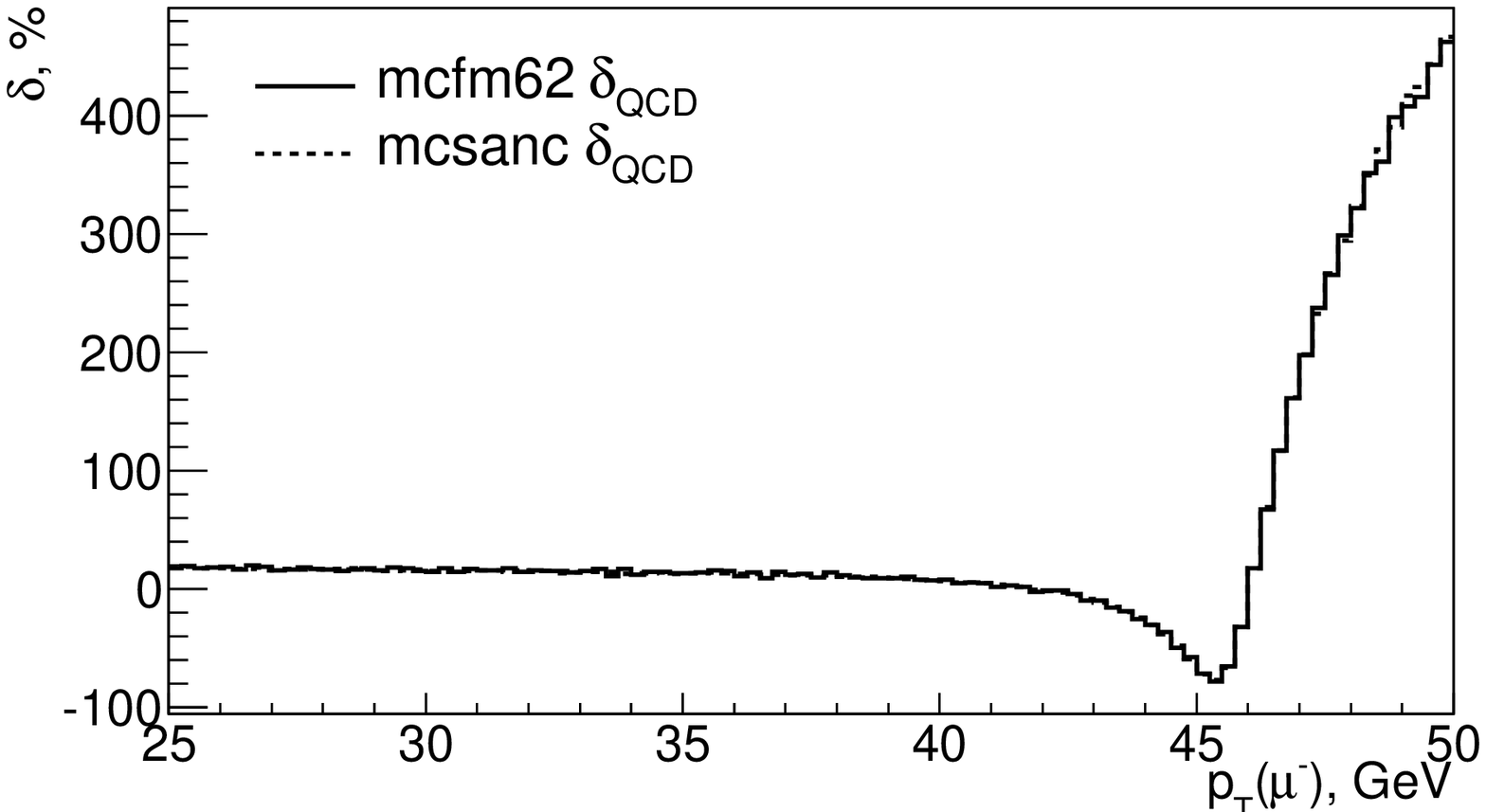}
\caption{Comparison of differential cross sections and correction factors
\(\delta(\mathrm{QCD})\) for neutral current Drell--Yan \(pp\to \mu^+\mu^-\)
process muon transverse momentum distributions.}
\label{fig:nc-vs-mcfm-pt4}
\end{figure}

Tables~\ref{tab:dy-total-xs}-\ref{tab:st-total-xs} contain results for
integrated LO and NLO EW and QCD cross sections obtained with the \texttt{mcsanc}-v1.01
integrator.  The LO and NLO QCD values are in agreement with the MCFM~\cite{Campbell:2010ff}
program within statistical error. A detailed comparison of differential neutral
current Drell--Yan cross section is shown on Figures~\ref{fig:nc-vs-mcfm-minv} 
and~\ref{fig:nc-vs-mcfm-pt4} for
dilepton invariant mass and lepton transverse momentum distributions correspondingly. The lower
plots on these figures show good agreement between NLO QCD correction factors obtained with
\texttt{mcsanc}-v1.01 and MCFM.

\begin{figure}[h]
  \includegraphics[width=0.5\textwidth]{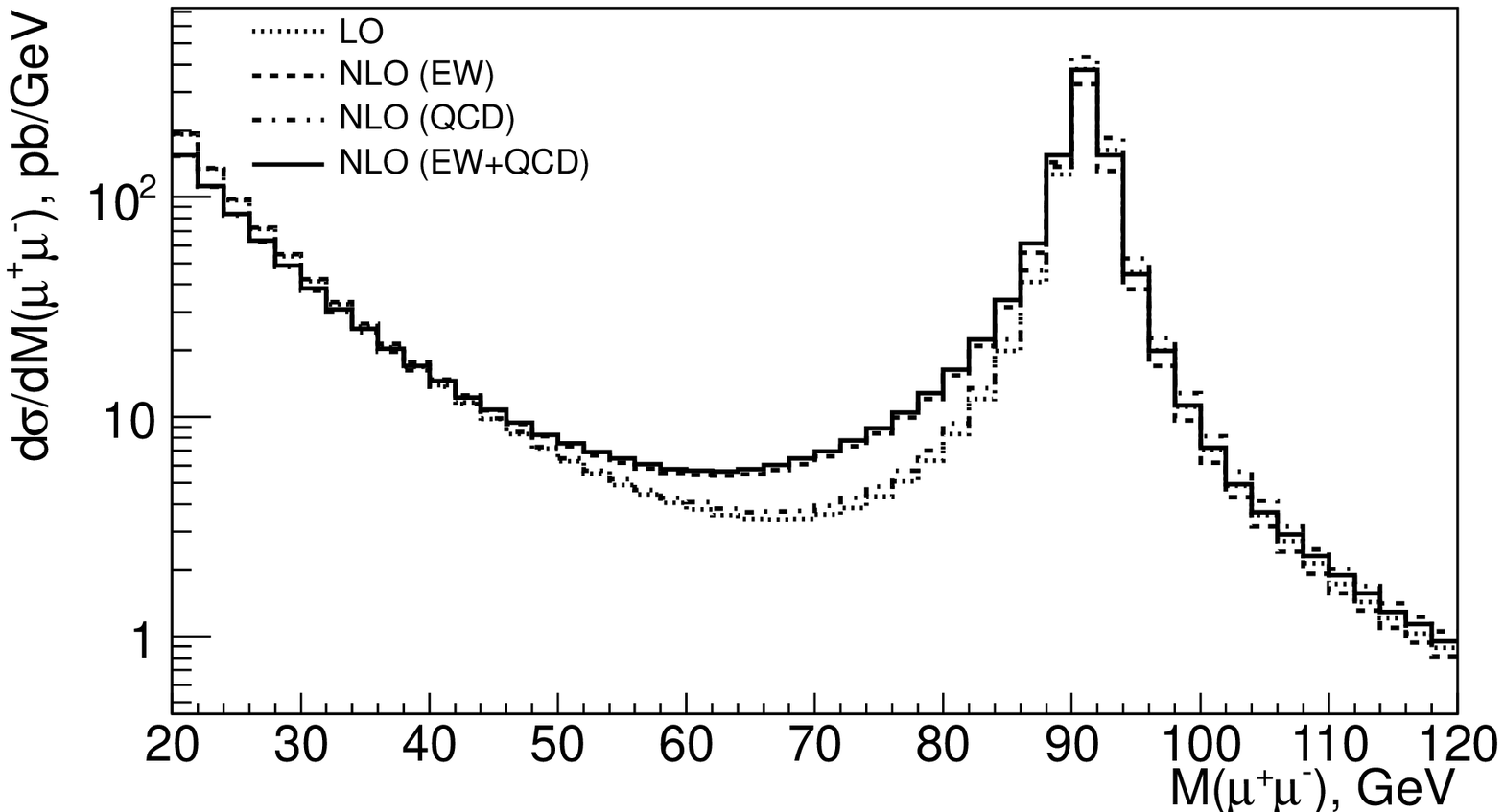}
  \includegraphics[width=0.5\textwidth]{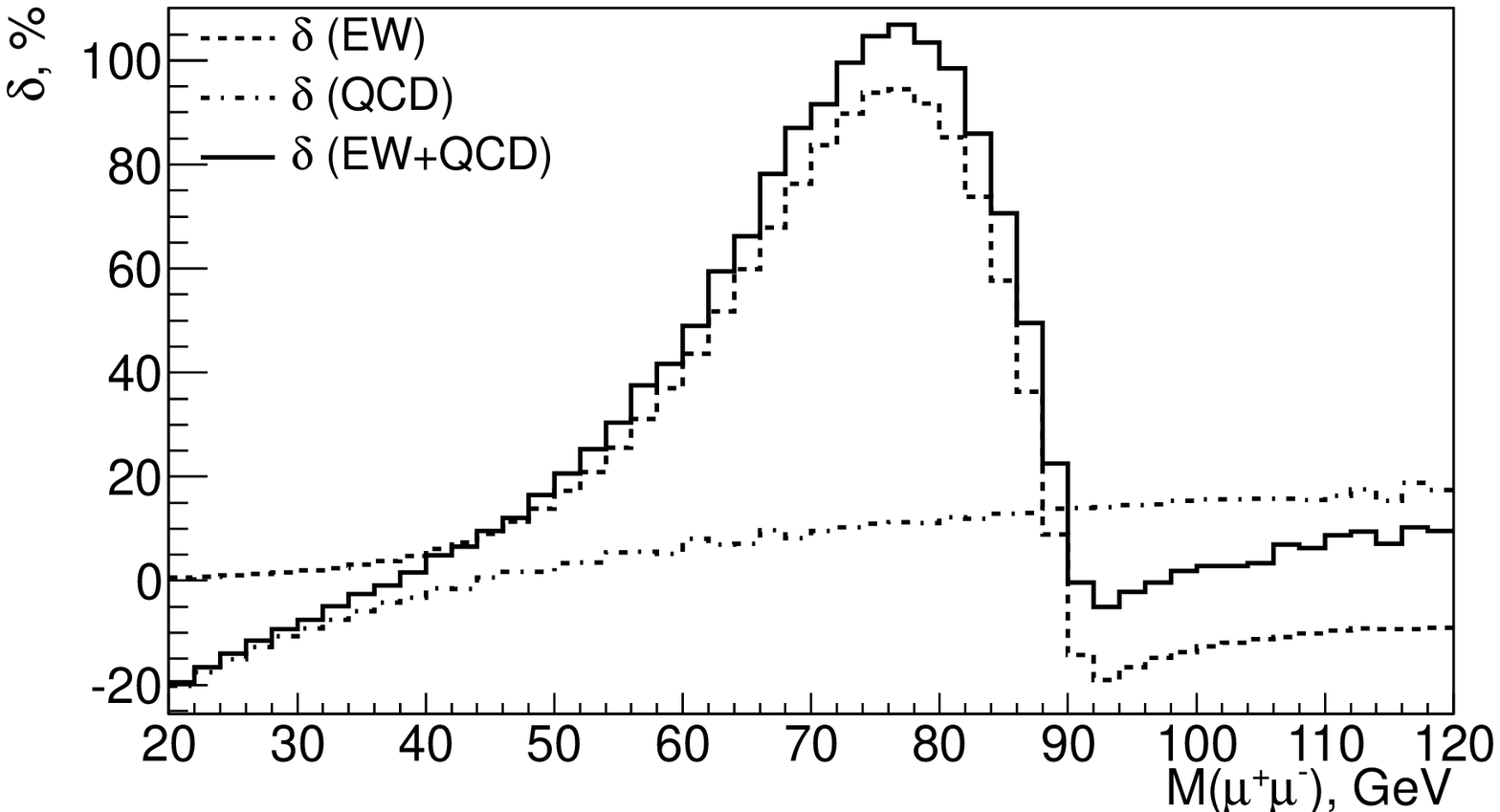}
\caption{A comparative representation of EW and QCD contributions to NLO
corrections and their sum for NC DY process \(pp\to \mu^+\mu^-\) for the dilepton
invariant mass distribution.}
\label{fig:nc-ew+qcd-minv}
\end{figure}

\begin{figure}[h]
  \includegraphics[width=0.5\textwidth]{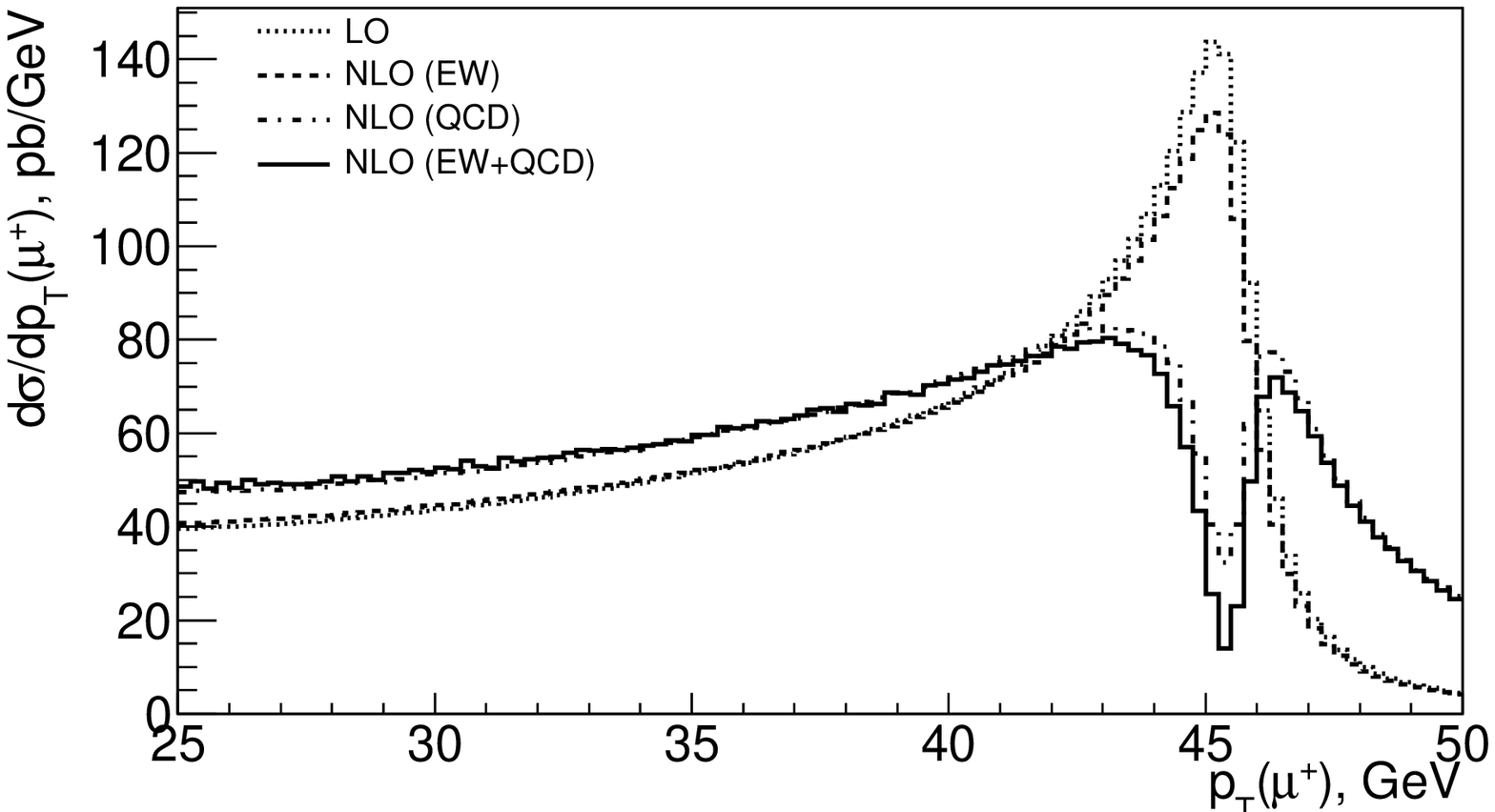}
  \includegraphics[width=0.5\textwidth]{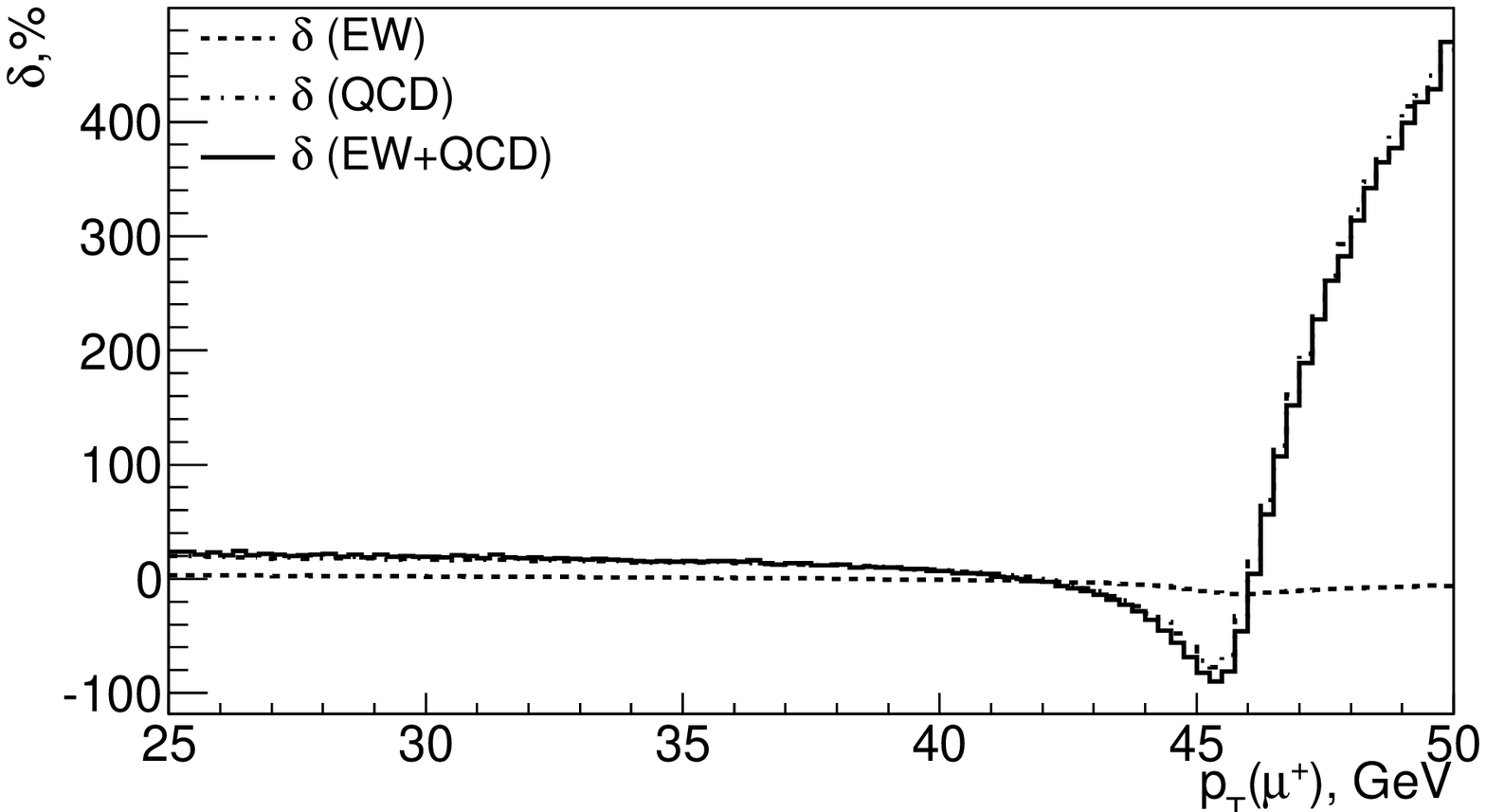}
\caption{A comparative representation of EW and QCD contributions to NLO
corrections and their sum for NC DY process \(pp\to \mu^+\mu^-\) for the lepton
transverse momentum distribution.}
\label{fig:nc-ew+qcd-pt3}
\end{figure}

Plots on Figures~\ref{fig:nc-ew+qcd-minv},~\ref{fig:nc-ew+qcd-pt3} show
separate EW and QCD contributions to the differential NC DY cross section and
their sum. The electroweak radiation produce high corrections around the Z
resonance peak for the dilepton invariant mass distribution, while the QCD
corrections are flatter.

\section{Summary \label{sec:summary}}
The presented \texttt{mcsanc} integrator is a new Monte Carlo tool for evaluating
higher order (NLO) EW and QCD cross sections. The integrator is based on the
SANC framework modules and calculates integrated and differential cross sections
for Drell--Yan processes, associated Higgs and gauge boson production and single-top quark
production. The code was thoroughly cross checked against another tools to 
provide consistent results. \texttt{mcsanc} uses advantage of multicore
implementation of the Cuba Monte Carlo integration library and supports
simple histogramming setup.

\section{Acknowledgements}
The authors are deeply thankful to D.~Bardin, L.~Kalinovskaya and other SANC group members for their help
and encouragement in developing the \texttt{mcsanc} integrator and writing
this paper. This work was supported in part 
by the RFBR grants 12-02-91526-CERN\_a and by the Dynasty Foundation.

\appendix

\section{mcsanc program \label{sec:mcsanc-prog}}

\subsection{Installation}
The LHAPDF~\cite{Whalley:2005nh} is required to be installed before starting
the installation.  After downloading and unpacking the tarball with
\texttt{mcsanc} from http://sanc.jinr.ru perform the following steps.
\begin{verbatim}
cd mcsanc_vXX 
autoreconf --force --install
./configure [--with-LHAPDF=/custom/LHAPDF/path]
make
\end{verbatim}

If succeeded the executable will be created in \texttt{./src} directory. The program
is launched with a command from \texttt{./share}:
\begin{verbatim}
cd ./share
../src/mcsanc [custom-input.cfg]
\end{verbatim}

Upon completion an output file \texttt{mcsanc}-\texttt{[run\_tag]}-\texttt{output.txt} with final results, run parameters and 
histograms will be created in the current directory.

\subsection{Configuration}
The \texttt{mcsanc} program reads to configuration files upon start: \texttt{input.cfg} and \texttt{ewparam.cfg}.
The first file, \texttt{input.cfg}, contains general steering parameters for a run,
VEGAS parameters, kinematic cuts and histogram parameters organized in Fortran namelists.
\subsubsection{Process namelist}
\begin{description}
\item [processId] defines a process to calculate (for available processes see Table \ref{tab:mcsanc-proc-list})(\texttt{integer}).
\item [run\_tag] is an arbitrary string value (\texttt{character*256}).
\item [sqs0] sets the beam energy at CMS frame (\texttt{double} \texttt{precision}).
\item [beams(2)] defines the incoming beams. So far only proton-proton beams are supported, which corresponds to \texttt{beams = 1,1} (\texttt{integer}).
\item [PDFSet] allows to set the parton density functions connected via LHAPDF (\texttt{character*256}).
\item [PDFMember] selects a member of PDF set. Usually 0 is used, which corresponds to central value (\texttt{integer}).
\item [iflew(5)] flags controlling electroweak components of the NLO EW computations. See below. (\texttt{integer}).
\begin{description}
\item [iqed]  = \hfill \\
  0: disables QED corrections\\
  1: with full QED corrections\\
  2: only initial state QED radiation (ISR)\\
  3: initial-final QED radiation interference term (IFI) \\
  4: only final state QED radiation (FSR)\\
  5: sum of initial and final state radiation contributions [IFI+FSR] (only for pid=\(001\div 003,\pm101\div 103\))\\
  6: sum of initial state and initial-final QED interference terms [ISR+IFI] (only for pid=\(001\div 003,\pm101\div 103\))
\item [iew] = 0/1 corresponds to disabled/enabled weak corrections.
\item [iborn] 0 or 1 selects respectively LO or NLO level of calculations.
\item [ifgg] = choice of calculations for photonic vacuum polarization \(\mathcal{F}_{\gamma\gamma}\):\hfill \\
 -1: 0 \\
  0: 1 \\
  1: \(1+\mathcal{F}_{\gamma\gamma}(\mathrm{NLO})\) \\
  2: \(1/[1 - \mathcal{F}_{\gamma\gamma}(\mathrm{NLO})]\)
\item [irun] = 0/1 corresponds to fixed/running gauge boson width
\end{description}
\item [iflqcd] controls if NLO QCD is calculated (\texttt{integer}).
\item [imsb] allows to select a subtraction scheme: 0/1/2 = none/MSbar/DIS (\texttt{integer}).
\item [irecomb] electron recombination option: 0/1 = off/on (\texttt{integer}).
\end{description}

\subsubsection{VegasPar namelist}
\texttt{VegasPar} regulates VEGAS parameters:
\begin{description}
\item [relAcc] relative accuracy limit on the calculated cross section (\texttt{double} \texttt{precision})
\item [absAcc] absolute accuracy limit on the cross section (\texttt{double} \texttt{precision})
\item [nstart] number of integrand evaluations performed at first iteration (\texttt{integer})
\item [nincrease] increment of integrand evaluations per iteration (\texttt{integer})
\item [nExplore] number of evaluations for grid exploration (\texttt{integer})
\item [maxEval] maximum number of evaluations (\texttt{integer})
\item [seed] random generator seed (\texttt{integer})
\item [flags] Cuba specific flags. See Cuba manual for more information~\cite{Hahn:2004fe} (\texttt{integer})
\end{description}

Different contributions to NLO cross sections (real, virtual) can differ from
each other by orders of magnitude. The combination of relative and absolute
accuracy limits gives a suitable way to control accuracy of total cross section
without need to limit each contribution. For example, the goal is 0.1\%
relative accuracy on total cross section. The user can set this value to the
\texttt{relAcc} option with an allowance for summing of the contributions. The
absolute error limit can then be calculated multiplying this value by the
largest estimated contribution. This will cut off unnecessary long calculations
for smaller contributions.

\subsubsection{KinCuts namelist}

This configuration section contains typical kinematic cuts for fiducial cross section calculation. Besides
original cuts, the user is free to implement his own in the \texttt{src/kin\_cuts.f}. The cut values
are set under corresponding name in the table in \texttt{input.cfg}

{\footnotesize
\begin{verbatim}
&KinCuts
  cutName       = 'm34',  'mtr',  'pt3', 'pt4',  ... 
  cutFlag       = 1,      0,      1,     1,      ...
  cutLow        = 66d0,   66d0,   20d0,  20d0,   ...
  cutUp         = 116d0,  116d0,  7d3,   7d3,    ...
/
\end{verbatim}
}
The \texttt{cutName} fields are fixed, \texttt{cutFlag} - 0/1 on/off cuts.
By default seven kinematic cuts are foreseen. They are: invariant \texttt{m34}
and transversal \texttt{mtr} masses, transversal momenta \texttt{pt3, pt4},
rapidities \texttt{eta3, eta4} of the final particles and \texttt{dR} parameter
to calculate recombined lepton momentum.

\subsubsection{Histograms with equidistant bins}

Namelist \texttt{FixedBinHist} allows to control histograms with fixed step. The histogram
parameters are defined in a table: under the histogram name, the user can set printing flag, 
min and max histogram values and a step. The binary printing flag controls if histogram is 
printed in the output (first bit) and if it's normalized to the bin width (second bit). 
For example to print unnormalized histograms set the flag to 1, to print normalized to 3.

{\footnotesize
\begin{verbatim}
&FixedBinHist
  fbh_name      = 'm34', 'mtr', 'pt34', 'pt3' 'pt4' ...
  fbh_flag      = 3,     3,     0,      3,    3,    ...
  fbh_low       = 66d0,  66d0,  20d0,   20d0, 20d0, ...
  fbh_up        = 116d0, 116d0, 58d0,   58d0, 58d0, ...
  fbh_step      = 2d0,   2d0,   1d0,    1d0,  1d0,  ...
/
\end{verbatim}
}

The most common parameters like invariant \texttt{m34} and transversal
\texttt{mtr} masses, transversal momenta \texttt{pt3, pt4}, rapidities
\texttt{et3, et4} of the final particles and transversal momenta
\texttt{pt34}, rapidity \texttt{et34} and \(\phi^*\) \texttt{phis} of
the intermediate boson are included by default. 

Any user defined histograms can be implemented in the \texttt{src/kin\_cuts.f} file. There 
the histogram names have no technical meaning, the requirement is that the order
of histograms in the input file must correspond to the histogram array \texttt{hist\_val} 
and the \texttt{nhist} variable set to the number of requested histograms.

\subsubsection{Histograms with variable bins}

Variable bin histograms are defined in namelist \texttt{VarBinHist}. The format is different
from \texttt{FixedBinHist}. The user have to set total number of variable step histograms
with \texttt{nvbh} parameter. Then follows definition of corresponding to \(i\)'th histogram name 
(\texttt{vbh\_name(i)}), printing flag (\texttt{vbh\_flag(i)}), 
number of bins (\texttt{vbh\_nbins(i)}) and array for the bin edges 
(\texttt{vbh\_bins(i,1:[nbins+1]}) as a space separated string of double precision numbers.
It is important that the variable bin histograms are available only with \texttt{gcc-v4.5.0} 
and higher.

{\footnotesize
\begin{verbatim}
&VarBinHist
  nvbh          = 7,

  vbh_name(1)	= 'm34',
  vbh_flag(1)	= 3,
  vbh_nbins(1)	= 6,
  vbh_bins(1,1:7) = 50d0 55d0 60d0 70d0 100d0 200d0,

  vbh_name(2)	= 'mtr',
  vbh_flag(2)	= 0,
  vbh_nbins(2)	= 6,
  vbh_bins(2,1:7) = 50d0 55d0 60d0 70d0 100d0 200d0,

  vbh_name(3)	= 'pt34',
  vbh_flag(3)	= 3,
  vbh_nbins(3)	= 5,
  vbh_bins(3,1:10) = 25d0 30d0 50d0 70d0 90d0 110d0,

  ...

/
\end{verbatim}
}

Some common kinematic variables are included in variable bin histograms by default.

\subsection{Electroweak parameters}

Electroweak parameters are set in ewparams.cfg configuration file. It contains \texttt{\&EWPars} 
name list with the following fields.
\begin{description}
\item [gfscheme] flag selects electroweak scheme in which the calculation is performed (\texttt{integer}).
  The possible values are 0:$\alpha(0)$-scheme, 
2:$G_{\mu}$-scheme, 4:$\alpha(M_Z)$-scheme (only for pid\(=001\div003\)).
\item [fscale] is the factorization scale (\texttt{double} \texttt{precision}). When equals -1, the value appropriate for
  requested process is set (e.g. final lepton invariant mass for DY process). Otherwise the given value is used.
\item [rscale] - renormalization scale defined with the same rules as \texttt{fscale} (\texttt{double} \texttt{precision}).
\item [ome] is the parameter separating contributions from soft and hard photon radiation (\texttt{double} \texttt{precision}).
\item [alpha, gf, alphas, conhc] - a list of constants and coefficients: 
  \(\alpha_{EM}\), \(G_{\mu}\), \(\alpha_{S}\), conversion constant (\texttt{double} \texttt{precision})
\item [mw, mz, mh] boson masses W, Z, Higgs (\texttt{double} \texttt{precision})
\item [wz, ww, wh, wtp] widths for bosons W, Z, Higgs and the top quark (\texttt{double} \texttt{precision})
\item [Vqq] CKM matrix (\texttt{double} \texttt{precision})
\item [men, mel, mmn, mmo, mtn, mta] leptonic masses \(\nu_e,e,\nu_{\mu}, \mu, \nu_{\tau}, \tau\)(\texttt{double} \texttt{precision})
\item [mdn, mup, mst, mch, mbt, mtp] quark masses \(u,d,s,c,b,t\)(\texttt{double} \texttt{precision})
\item [rmf1] is a 12 element array of fermion masses for technical purposes (\texttt{double} \texttt{precision})
\end{description}

Vegas grids have to be generated from scratch if these parameters are changed.

\subsection{Persistency}
Cuba library \cite{Hahn:2004fe} allows to save the VEGAS grid for further restoration.
The integration stops when requested accuracy is reached. The last version of the grid and 
histograms state are saved in \texttt{*.vgrid} and \texttt{*.hist} files. If the user desires
to improve precision he can set the new limits in \texttt{input.cfg}
and restart the program in the directory with saved grids. Care should be 
taken that all other configurations (PDF, electroweak parameters) are remained
the same. If the user wishes to change the histograms he can use 
old grids, but remove \texttt{*.hist} files and fill the histograms from scratch.
The histogram errors will correspond to less precision than the total cross 
section.

The grid exploration stage is omitted if grid state file is found.

\subsection{Output}

The main output of the program will be written to
\texttt{mcsanc}-\texttt{[run\_tag]}-\texttt{output.txt} file. The file contains
final table with cross sections from the components and the summed total cross
section. After the table follows the CPU usage, and a list of input parameters,
followed by a list of cuts applied in the process of integration. The remaining
of the output file is a section containing Histograms listing.  Each histogram
is output separately in a text file for plotting convenience. The file name
format is \texttt{[run\_tag]}\_\texttt{[hist\_name]-XXX}, where \texttt{XXX}
stands for \texttt{born, tota} or \texttt{delt} for LO, NLO and \(\delta{\%}\)
histograms.

\section{Multiprocess calculations\label{sec:multiproc-eff}}
The Cuba library \cite{Hahn:2004fe}, used as a Monte Carlo integrating tool in the \texttt{mcsanc}
program, supports multiprocessing. Having a multicore CPU the user can request
the program to run on several cores by specifying \mbox{\texttt{CUBACORES}} environment
variable.

\begin{figure}[h]
\begin{center}
\includegraphics*[width=0.4\textwidth]{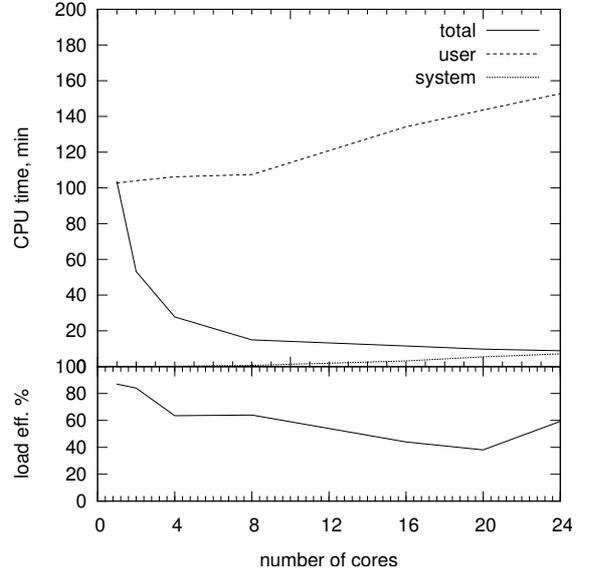}
\caption{CPU usage and load efficiency for \texttt{mcsanc} program depending
on the number of processor cores.}
\label{fig:cpu-usage}
\end{center}
\end{figure}

Due to the features of inter-process communications, the efficiency of
multiprocessing is not always optimal. Figure~\ref{fig:cpu-usage} shows NLO EW
cross section calculation time depending on the number of CPU cores active. The
test was run on a dual-processor \mbox{Intel\textsuperscript{\textregistered}}
\mbox{Xeon\textsuperscript{\textregistered}} machine with 12 real (24 virtual)
cores with Linux operating system.

The upper plot summarizes multicore CPU productivity: ``total'' is the
wall clock time passed during the run; ``user'' is the CPU time consumed by the
program (roughly equals to wall clock time multiplied by the number of cores in
case of 100\% efficiency); ``system'' is the time spent by the operating system
on the multiprocessing service.  One can see that the multiprocessing is
efficient with number of cores up to 8, after which the total run time does not
significantly decrease and the CPU time (``user'') grows.  It is also clear from
the lower plot that the average CPU load efficiency drops below 50\% with more than 8
cores active.
\bibliographystyle{utphys_spires}
\bibliography{MCSANC}

\end{document}